\documentstyle[12pt]{article}
\begin{document}
%
\newcommand{\beq}{\begin{equation} }
\newcommand{\eeq}{\end{equation} }
\newtheorem{theorem}{Theorem}
\newtheorem{lemma}{Lemma}
\newtheorem{propo}{Proposition}
\newtheorem{defin}{Definition}
\newenvironment{coroll}{\begin{quotation}{\bf Corollary:\/}}{\end{quotation} }
\newenvironment{proof}{{\em Proof:}\/}{$\Box$ \medskip \\}
%
%
\newcommand{\eps}{\varepsilon}
\renewcommand{\star}{\ast}
\newcommand{\id}{\mbox{\rm id}}

\title{Quantum Principal Fiber Bundles:\\
Topological Aspects
\thanks{Supported by KBN under grant 2 1047 91 01} }
\author{Robert J. Budzy\'{n}ski
\thanks{email: budzynsk@fuw.edu.pl}\\
Institute of Theoretical Physics\\
Warsaw University
\and
Witold Kondracki
\thanks{email: witekkon@impan.gov.pl}\\
Institute of Mathematics\\
Polish Academy of Sciences}
\date{IM PAN 517 \\ hep-th/9401019 \\ December 1993}
\maketitle
\begin{abstract}
We introduce the notion of locally trivial quantum principal bundles.
The base space and total space are compact quantum spaces
(unital $C^{\star}$-algebras), the structure group is a compact
matrix quantum group.
We prove that a quantum bundle admits sections if and only if it is
trivial. Using a quantum version of \v{C}ech cocycles, we obtain a
reconstruction theorem for quantum principal bundles. The classification
of bundles over a given quantum space as a base space is reduced
to the corresponding problem, but with an ordinary classical group
playing the role of structure group. Some explicit examples are
considered.
\end{abstract}
\section{Introduction}
Quantum groups \cite{woron1,rtf,drin} are by now a well-established
notion, and one that attracts considerable interest, both as a subject of pure
mathematics and as a tool in theoretical physics. One of the approaches
to quantum groups is to view them within the wider context of
non-commutative geometry \cite{con}, as quantum spaces endowed
with a particular additional structure. More specifically, we adopt
here the point of view, developed by Woronowicz and collaborators
\cite{woron1,podles1,podles2,woron2}, where a (generally non-commutative)
$C^{\star}$-algebra is interpreted as a generalization of the
(commutative) $C^{\star}$-algebra of continuous functions on a
locally compact topological space. In this sense, the theory of
$C^{\star}$-algebras may be considered as an extension of the
theory of a certain category of classical topological spaces
(point sets)\footnote{For an example of work along these lines,
see \cite{Klimek}. }.
Such a framework provides a convenient starting point for further
development of non-commutative geometry: classical geometrical notions,
such as differential structures, differential forms, metrics, connections,
etc. are to be re-defined in a way applicable to non-commutative
$C^{\star}$-algebras.

Within this context, it is natural to seek non-commutative (or `quantum')
extensions of classical geometrical constructions involving Lie
groups, now generalized to matrix quantum groups \cite{woron1}.
For instance, quantum homogeneous spaces and quotient spaces of
quantum groups were studied by Podle\'s \cite{podles1,podlesPhD,podles2}.
The aim of the present paper is to lay the groundwork for a theory
of locally trivial quantum principal fiber bundles. Such a construction
is of obvious intrinsic interest, as it generalizes a very important
and natural object of classical geometry. On the physical side, it
is well known that principal bundles provide the natural geometrical
setting for classical Yang-Mills theory. One may expect that quantum
principal bundles should play a corresponding role for a non-commutative
Yang-Mills theory, attempts to formulate which are currently under way.

In the present work, we follow the general approach to quantum spaces
sketched above: the total bundle space and base space are replaced by
(compact) quantum spaces, represented by unital $C^{\star}$-algebras,
and the structure group is replaced by a (compact matrix) quantum group.
The classical definitions are extended by translating them into dual
form, involving the algebras of continuous functions on the corresponding
spaces, and giving up commutativity. This is not entirely straightforward:
the quantum definitions and statements must make no reference to points
of the spaces involved (quantum spaces are in no sense point sets).
An example of the difficulties one encounters is given by the concept
of free action of a group. This does not appear to admit a satisfactory
generalization to quantum spaces. For instance, the definition proposed
in  \cite{dur,maj} is formulated in terms of a mapping which is not
a $C^{\star}$-algebra morphism. Our choice is to impose a quantum
version of local triviality, a stronger requirement.

The scope of the present paper is restricted to formulating
the basic definitions and extracting their most immediate consequences,
specifically, those of a `topological' nature. The study of differential
geometric structures on quantum principal bundles within the framework
of our proposed theory is left to future publications. Moreover, it
should be pointed out that our work can be extended in a number of
directions. For instance, one may work with open (instead of closed)
coverings of the base space of the bundle, as is more conventional
in classical geometry; this is a rather technical point, involving
the more intricate theory of noncompact quantum spaces \cite{woron2}.
A more substantial point concerns our choice of the quantum extension of
the notion of Cartesian product; though certainly not a unique choice,
it is the one most frequently considered in similar work \cite{woronpc}.

In section 2 we review the basic definitions and facts concerning
quantum group theory as developed by Woronowicz. Section 3 introduces
trivial quantum principal bundles, their sections and automorphisms.
Bundle automorphisms are defined in such a way that their set forms
a group in the ordinary sense. In the commutative case, this group
coincides with the usual group of gauge transformations of a trivial
principal bundle. In addition, the classical one to one correspondence
between sections and trivializations also extends to our theory.
Section 4 begins with the general definition of locally trivial
quantum principal bundles. Our definition is somewhat more restrictive
than the one proposed in \cite{maj}. This is because we were careful to
preserve in full the correspondence with the classical theory: if all
`function' algebras are taken to be commutative, all the objects of
our theory reduce to their classical counterparts. Such an approach
allows us to prove a number of powerful results; the first
(Th. \ref{sections}) states that existence of a section is equivalent
to global triviality of the bundle. Next, we present an outline
of \v{C}ech cohomology theory for quantum principal bundles,
which is then employed to give a reconstruction theorem
(Th. \ref{reconstruction}), stating that a quantum
principal bundle may be recovered from the corresponding quantum
\v{C}ech cocycle. The next of our main results (Th. \ref{subb})
is that two bundles with isomorphic sub-bundles are themselves
isomorphic. Theorem \ref{csb} states that every bundle has what
we call a `classical sub-bundle', {\em i.e.} a sub-bundle with
the structure group being an ordinary (classical) group.
Taken together with Theorem \ref{classic}, the classification of
quantum principal bundles is reduced to that
of bundles over the same base space, with the structure group
being the classical subgroup of the corresponding quantum group.
This extends considerably a result of \cite{mul}, where
only bundles over commutative base spaces were considered.

Following a brief digression on associated bundles in section 5,
in section 6 we present some explicit examples: we describe all
$SU_q(2)$ principal bundles over the quantum unit disk of Klimek
and Lesniewski \cite{Klimek}, and over the Podle\'s spheres
\cite{podles2}. An appendix is devoted to a brief summary of
the principal concepts and operations of the theory of quantum
spaces.

\section{Quantum groups}
All $C^{\star}$-algebras which will be considered in this paper are
understood to be separable $C^{\star}$-algebras with unit; correspondingly,
all algebra homomorphisms are unital $C^{\star}$-algebra homomorphisms.
Homomorphisms (linear, multiplicative, $\ast$-preserving mapppings)
of $C^{\star}$-algebras into the complex number field
will be termed functionals for the sake of brevity.
We use the notation $m_A : A\otimes A \rightarrow A$ (where $A$ is
a $C^{\star}$-algebra) for the linear
mapping (not an algebra homomorphism in general) defined by
$m_A (a\otimes b)= ab$.
For a review of the basic notions of the theory of compact quantum spaces,
as they are applied in the present paper, we refer the reader to the
appendix.

The definition below follows \cite{woron1}.
\begin{defin}
$G = (A,u)$ is called a (compact matrix) quantum group if $A\ne \{0\}$
is a $C^{\star}$-algebra, $u = (u_{ij})$ is a $N \times N$ matrix with
entries in $A$, and
\begin{enumerate}
\item{$A$ is the smallest $C^{\star}$ algebra containing all
matrix elements of $u$,}
\item{There exists a homomorphism $\Delta :A \rightarrow A\otimes A$ such that
\beq
\Delta(u_{ij}) = \sum_k u_{ik} \otimes u_{kj},
\eeq
}
\item{$u$ and $u^T$ are invertible.}
\end{enumerate}
\end{defin}
In the case when $A$ is a commutative algebra, $(A,u)$ will be called
a classical (matrix) group. We will often employ the notation $A=C(G)$.

The above definition implies the following
\begin{propo}
\begin{enumerate}
\item $(\Delta \otimes \id )\Delta = (\id \otimes \Delta )\Delta$
(co-associativity);
\item There exists a unique functional $\eps$ on $A$, the counit,
with the properties
\beq
\eps (u_{ij}) = \delta _{ij}
\eeq
\beq
\label{counit}
(\eps \otimes \id )\Delta = (\id \otimes \eps )\Delta = \id .
\eeq
\item Let us denote by ${\cal A}$ the dense
subalgebra in $A$ generated by $u_{ij}$.
There exists a (unique) anti-homomorphism (i.e. linear and anti-multiplicative
mapping) $S : {\cal A} \rightarrow {\cal A}$ called the antipode, such that
$$S(u_{ij}) = (u^{-1})_{ij}$$
$$S(S(a^{\star})^{\star}) = a$$ for all $a \in {\cal A}$.
%
%
\end{enumerate}
\end{propo}
A {\em representation} of the quantum group $G$ is an invertible $M \times
M$ matrix $v$ with entries in $A$ such that $\Delta (v_{mn}) = \sum_p
v_{mp}\otimes v_{pn}$. In particular, $u$ is a representation of $G$.
For any representation $v$ of $G$, $\eps (v_{mn}) = \delta _{mn}$ and
$S(v_{mn}) = (v^{-1})_{mn}$.

Two representations are called equivalent if the two corresponding matrices
are related by a similarity transformation given by a matrix with
entries in ${\bf C}$. A representation is reducible if its matrix is (up to
similarity) block-diagonal. Otherwise, a representation is called
irreducible. For more details, see \cite{woron1}.

\begin{defin}
A subgroup of the quantum group $G=(A,u)$ is the triple $(G,H,\theta_{HG})$,
where $H=(B,v)$ is a quantum group, $\dim v = \dim u$, and
$\theta _{HG}: A \rightarrow B$ is a $C^{\star}$ homomorphism such that
$\theta _{HG}(u_{ij}) = v_{ij}$.
\end{defin}
Note that $\theta _{HG}$ is necessarily a $C^{\star}$ epimorphism.

We now introduce a special subgroup, which exists for any quantum group
$G$, and which will play an important role in the sequel. Moreover,
this {\em classical subgroup} corresponds to a group in the usual sense.

For a quantum group $G$ consider the set of functionals on $C(G)$. This
set $G/$ is equipped with the natural structure of a group: the group
multiplication is given by
$$\phi \cdot \psi = (\phi \otimes \psi)\Delta$$
for $\phi, \psi \in G/$; the neutral element is the functional $\eps$,
the co-unit. We thus see that $G/$ is a semigroup with unit; but, since
it is also a compact topological space, $G/$ is a group \cite{woron1}.
In fact, it is a compact subgroup of $GL(N,{\bf C})$ with $N = \dim u$:
its elements are given
by $\phi (u_{ij})$ for $\phi \in G/$, and group multiplication is
equivalent to matrix multiplication.

Consider now $C(G/)$, the algebra of continuous functions on the group
$G/$. One can define a natural homomorphism $\rho : C(G) \rightarrow
C(G/)$ by the formula
$$[\rho (f)](\phi) = \phi (f).$$
The triple $(G, G/, \rho)$, where $G/$ as a quantum group is $G/ =
(C(G/),\rho(u_{ij}))$, is what we call the classical subgroup of $G$.

Observe that the kernel of $\rho$ coincides with the {\em commutator}
of $C(G)$, i.e. the smallest closed $\star$-ideal in $C(G)$ which
contains the commutator of any two elements of $C(G)$.
\begin{defin}
Let $C(X)$ be a $C^{\star}$ algebra with unit, and $G$ be a quantum group.
We say that a $C^{\star}$ homomorphism $\Gamma : C(X) \rightarrow
C(X) \otimes C(G)$ is an action of $G$ on the quantum space $X$ if:
\begin{enumerate}
\item $(\Gamma \otimes \id)\Gamma = (\id \otimes \Delta)\Gamma$,
\item The closure of the span of $(I \otimes y)\Gamma x,\: x \in C(X),\:
y \in C(G)$ is equal to $C(X) \otimes C(G)$.
\end{enumerate}
\end{defin}
The above definition is taken from \cite{podles1}. For completeness, we
also quote the following theorem \cite{podles1} (see also \cite{podlesPhD}):
\begin{theorem}
\label{decomp}
Let $\Gamma$ be an action of the quantum group $G$ on the quantum space $X$.
Then $C(X)$ may be decomposed as the (closure of the) direct sum of invariant
subspaces corresponding to the inequivalent irreducible representations
of $G$, with the multiplicity (possibly infinite) of any given representation
being uniquely determined.
\end{theorem}

\section{Trivial quantum principal bundles}
In the present section we collect the basic facts concerning trivial
quantum principal bundles. This will serve to introduce the concepts
and methods which will be applied in the sequel to the general case
of (locally trivial) quantum principal bundles.
\begin{defin}
Let $C(P)$, $C(X)$ be $C^{\star}$ algebras, $G= (C(G),u)$ a quantum group,
$\Gamma : C(P) \rightarrow C(P) \otimes C(G)$ an action of $G$,
and $\pi : C(X) \rightarrow C(P)$ an injective homomorphism, such that
\beq
\Gamma \pi (f) = \pi (f) \otimes I
\eeq
for all $f \in C(X)$.

$(C(P), C(X), G, \pi , \Gamma)$ will be called a trivial quantum
principal bundle if there exists a $(C^{\star})$ isomorphism
$\Phi : C(P) \rightarrow C(X) \otimes C(G)$ such that
\beq
\Phi \pi (f) = f \otimes I
\eeq
for all $f \in C(X)$, and
\beq
(\Phi \otimes \id ) \Gamma = (\id \otimes \Delta) \Phi .
\eeq
 $\Phi$ will be called a trivialization of the bundle.
\end{defin}

\begin{defin}
Let $(C(P), C(X), G, \pi, \Gamma)$ and $(C(P'), C(X), G, \pi ',
\Gamma ')$ be trivial quantum principal bundles.

A $C^{\star}$ isomorphism $\Xi : C(P) \rightarrow C(P')$ will be
called an isomorphism of trivial quantum principal bundles if:
\beq
\Xi \pi = \pi '
\eeq
and
\beq
(\Xi \otimes \id ) \Gamma = \Gamma ' \Xi .
\eeq
\end{defin}

It therefore follows that a trivial quantum principal bundle is
isomorphic as a bundle to $(C(X)\otimes C(G), C(X), G, \id \otimes I,
\id \otimes \Delta)$, and the freedom to choose a trivialization
corresponds to automorphisms of this (product) bundle.

The following lemma provides a more explicit description of the
automorphisms of a trivial quantum principal bundle. It is the
main technical tool which will be employed in this paper.
\begin{lemma}
\label{tau-lemma}
Let $\Psi$ be an automorphism of the trivial principal bundle
$(C(X)\otimes C(G), C(X),G, \id~\otimes~I, \id~\otimes~\Delta)$.
The homomorphism
$\tau_{\Psi} : C(G) \rightarrow C(X)$, uniquely determined from $\Psi$
through the formula
\beq
\tau_{\psi} = (\id \otimes \eps )\Psi (I \otimes \id )
\eeq
takes values in the center of $C(X)$. Conversely, given any $\tau$
with the above property, the formula
\beq
\Psi_{\tau} = (m_{C(X)}\otimes \id )(\id \otimes \tau \otimes \id )
(\id \otimes \Delta )
\eeq
uniquely defines $\Psi_{\tau}$, which is a bundle automorphism.
\end{lemma}
\begin{proof}
The algebra $C(X) \otimes C(G)$ is generated by two mutually commuting
subalgebras, $C(X) \otimes I$ and $I \otimes C(G)$. The latter is in turn
generated by the elements $I \otimes u_{ij}$. Thus the automorphism
$\Psi$ is uniquely determined by the formulas
$$\Psi (f \otimes I) = f \otimes I,$$
$$\Psi (I \otimes u_{ij}) = \Psi_{ij} \in C(X) \otimes C(G).$$
Since $\Psi$ is an isomorphism,
\beq
\label{commute}
(f \otimes I) \Psi_{ij} = \Psi_{ij} (f \otimes I).
\eeq
The condition that $\Psi$ commutes with $\id \otimes \Delta$ leads to
$$(\Psi \otimes \id)(\id \otimes \Delta) I \otimes u_{ij} =
(\id \otimes \Delta)\Psi(I \otimes u_{ij}), $$
i.e.
$$(\Psi \otimes \id) I \otimes u_{ik} \otimes u_{kj} =
(\id \otimes \Delta)\Psi_{ij},$$
$$\sum_k\Psi_{ik}\otimes u_{kj} = (\id \otimes \Delta)\Psi_{ij}.$$
Applying to both sides $\id \otimes \eps \otimes \id$ we obtain
$$[(\id \otimes \eps)\Psi_{ik}]\otimes u_{kj} = \Psi_{ij}$$
by eq. \ref{counit}.

But $(\id \otimes \eps)\Psi_{ik} = \tau(u_{ik})$, therefore
$$\Psi_{ij} = \tau(u_{ik})\otimes u_{kj}.$$
Thus in virtue of eq. \ref{commute}:
$$(\tau (u_{ik})f - f\tau (u_{ik}))\otimes u_{kj} = 0.$$
Now, the matrix $u$ is invertible, hence $I\otimes u_{kj}$ are also
elements of an invertible matrix. It follows that
$$\tau (u_{ik})f - f\tau (u_{ik}) = 0$$
for all $f \in C(X)$.

This proves the first claim of the lemma.

Secondly, given $\tau : C(G)\rightarrow Z(C(X))$, ~$\Psi_{\tau}$ is a
homomorphism uniquely specified by the given formula. The presence of
the diagonal mapping $m : C(X)\otimes C(X) \rightarrow C(X)$ causes
no problems due to the values of $\tau$ being central.

It remains to be shown that $\Psi$ is an isomorphism, i.e. the inverse
$\Psi^{-1}$ exists. But it is easily verified that $\Psi^{-1}$ is
obtained in the analogous way from $\tau S$.
\end{proof}

A simple consequence of the above lemma is the folowing:
\begin{coroll}
Since $\tau$ takes values in the center of $C(X)$, $\tau$ vanishes
on the kernel of the projection onto the classical subgroup
$\rho : C(G) \rightarrow C(G/)$.
\end{coroll}
Thus every such $\tau$ is in one-to-one correspondence with a
homomorphism $\tau / :C(G/) \rightarrow Z(C(X))$, such that
$ \tau = \tau / \circ \pi / $.
\begin{propo}
The set of homomorphisms $\tau : C(G) \rightarrow Z(C(X))$ forms
a group isomorphic to the group of automorphisms of the trivial principal
bundle
$(C(X)\otimes~C(G), C(X), G, \id~\otimes~I, \id~\otimes~\Delta)$.
The group structure is given by
$$\tau_1 \cdot \tau_2 = m_{C(X)}(\tau_1\otimes \tau_2)\Delta$$
$$\tau^{-1} = \tau S$$
$$e = I_{C(X)}\eps$$
giving the multiplication, inversion, and unit element, respectively.
\end{propo}

\subsection{Sections of a trivial quantum principal bundle}
\begin{defin}
For a trivial principal bundle $(C(P), C(X), G, \pi, \Gamma)$,
let $s: C(P) \rightarrow C(X)$ be a homomorphism such that
$s \pi = \id$. We will call $s$ a section of the bundle.
\end{defin}

It turns out that, similarly as in the classical case, every section
of a trivial quantum principal bundle determines a trivialization,
and every pair of sections determines a bundle automorphism. This
is the subject of the following lemma.

\begin{lemma}
a) There exists a canonical one-to-one correspondence between
the set of sections of $(C(P), C(X), G, \pi, \Gamma)$ and the set
of trivializations.

b) Every pair of sections $s_1,\: s_2$ determines a unique bundle
automorphism $\Xi_{12} : C(P) \rightarrow C(P)$ such that
$s_1 \Xi_{12} = s_2$.
\end{lemma}
\begin{proof}
First let us observe that any trivialization $\Phi$ may be used
to determine a section, via
$$s_{\Phi} = (\id \otimes \eps )\Phi.$$
Observe now that given a section
$s' : C(X) \otimes C(G) \rightarrow C(X)$ of the product bundle,
an automorphism of the product bundle is obtained by
$$\Psi_{s'} = (s' \otimes \id )(\id \otimes \Delta ).$$
That this is indeed an automorphism is verified by observing that
$\tau_{s'}: C(G) \rightarrow C(X)$, given by
$$\tau_{s'} = s' (I \otimes \id),$$
satisfies the assumptions of lemma
\ref{tau-lemma}, and by the procedure of that lemma coresponds precisely
to $\Psi_{s'}$.

Next, observe that composing the section $s'$ with an automorphism
$\Xi : C(X) \otimes C(G) \rightarrow C(X) \otimes C(G)$ gives
$$\Psi_{s' \Xi} = \Psi_{s'}\Xi.$$
Taking an arbitrary
trivialization $\Phi : C(P) \rightarrow C(X) \otimes C(G)$, we
define $\Phi_s : C(P) \rightarrow C(X) \otimes C(G)$ by
$\Phi_s = \Psi_{s'}\Phi$, where $s' = s \Phi^{-1}$.
It is easily verified that $\Phi_s$ is independent of the choice
of $\Phi$. Moreover, for any trivialization $\Phi '$,
$$\Phi_{s_{\Phi '}} = \Phi '.$$
This establishes point a).

Now, take a pair of sections $s_1, s_2$. As a simple consequence
of the above, we can write
$$\Xi_{12} = \Phi_{s_1}^{-1}\Phi_{s_2} : C(P) \rightarrow C(P),$$
which fulfills point b), completing the proof.
\end{proof}

\section{Quantum principal bundles: general definition and basic
properties}
The present section begins with a general definition of (locally
trivial) quantum principal bundles. This definition is basically
a transcription, into the dual language of function algebras (now
not necessarily commutative) of the usual classical definition,
except for one feature: local triviality is prescribed by using
a finite covering of the base space by closed (not open) sets.
It is of course possible to use open coverings, but at the cost
of complications related to the more difficult theory of
{\em noncompact} quantum spaces (see e.g. \cite{woron2}).
As the present approach is
sufficient for the purposes addressed in this paper, we choose to
leave this extension to future work.
\begin{defin}
Let $C(P), C(X)$ be $C^{\star}$ algebras, $G = (C(G), u)$ a quantum
group, $\pi : C(X) \rightarrow C(P)$ an injective homomorphism,
$\Gamma : C(P) \rightarrow C(P) \otimes C(G)$ an action of $G$.

We will call $(C(P), C(X), G, \pi, \Gamma)$ a quantum principal bundle
if:

a) $\Gamma \pi = \pi \otimes I$

b) there exists a finite covering $(C(U_i), \kappa_i)_{i \in I}$ of
$C(X)$ and a family $(\hat{\kappa_i})_{i \in I}$ of surjective
homomorphisms $\hat{\kappa_i} : C(P) \rightarrow C(U_i)\otimes C(G)$
forming a covering of $C(P)$ and obeying the following equations:
$$\hat{\kappa_i} \pi = (\id \otimes I) \kappa_i, $$
$$(\hat{\kappa_i} \otimes \id ) \Gamma = (\id \otimes \Delta )\hat{\kappa_i}.$$
\end{defin}
Obviously, a trivial quantum principal bundle is a special case of
the present definition: take $\kappa = \id_{C(X)},\: \hat{\kappa} = \Phi$ ---
any trivialization. The definitions of bundle isomorphism and section
can be trivially extended to the general case.
\begin{theorem}
\label{sections}
A quantum principal bundle admits a section iff it is trivial.
\end{theorem}
\begin{proof}
By definition, a trivial bundle admits a trivialization, and it
was proven above that any trivialization determines a section.

To prove the converse: take the mapping
$$\Phi_s : C(P) \rightarrow C(X) \otimes C(G)$$
given by $\Phi_s = (s \otimes \id)\Gamma$. We claim that $\Phi_s$
is a trivialization for any section $s$.

It is easily verified that $\Phi_s \pi = \id \otimes I$ and
$(\Phi_s \otimes \id)\Gamma = (\id \otimes \Delta)\Phi_s$. It remains
to be shown that $\Phi_s$ is bijective.

First, observe that given a section $s$ and a family of local
trivializations $\hat{\kappa_i}$ over a covering $\kappa_i$, the section
$s$ descends to a family of local sections $s_i : C(U_i)\otimes C(G)
\rightarrow C(U_i)$, fulfilling $\kappa_i s = s_i \hat{\kappa_i}$ and
$s_i (\id \otimes I) = \id$. Indeed, since $\hat{\kappa_i}$ are surjective,
given (for fixed $i$)
$$C(U_i) \otimes C(G) \ni f_i = \hat{\kappa_i}(f)$$
the element $f \in C(P)$ is determined up to an element of
$\ker \hat{\kappa_i}$.
For $s_i$ to be well defined, we must show that
$\kappa_i s(\ker \hat{\kappa_i}) = \{0\}$.

Take $h \in \ker \hat{\kappa_i}$; $h$ may be uniquely decomposed
into
$$h = \pi s(h) + (h - \pi s(h)),$$
where the second term is in $\ker s$, in virtue of $s\pi = \id$. One has
therefore
$$\hat{\kappa_i}(h) = 0 =\hat{\kappa_i}\pi s(h) =
(\id \otimes I)\kappa_i s(h),$$
hence $\kappa_i s(h) = 0$, as required.

Next, the family of sections $s_i$ determines a family of automorphisms
$$\Phi_{s_i} :C(U_i)\otimes C(G) \rightarrow C(U_i)\otimes C(G),$$
$$\Phi_{s_i} = (s_i \otimes \id )(\id \otimes \Delta ).$$
One easily verifies that
\beq
\label{Phis}
\Phi_{s_i}\hat{\kappa}_i = (\kappa_i \otimes \id )\Phi_s.
\eeq
Consider now the direct sum algebra $\bigoplus_{i} C(U_i)\otimes C(G)$,
the direct sum of local automorphisms $\bigoplus_{i} \Phi_{s_i}$, and
the morphisms
$$\hat{\cal K} : C(P) \rightarrow \bigoplus_{i} C(U_i)\otimes C(G),$$
$${\cal K}\otimes \id : C(X)\otimes C(G) \rightarrow \bigoplus_{i}
C(U_i)\otimes C(G),$$
defined by
$$\hat{\cal K} = \bigoplus_{i} \hat{\kappa}_i,$$
$${\cal K}\otimes \id = \bigoplus_{i}(\kappa_i \otimes \id).$$
Since $\bigcap_i \ker \kappa_i = \{ 0 \}$ and $\bigcap_i \ker
\hat{\kappa}_i = \{ 0 \}$, $\hat{\cal K}$ and ${\cal K}\otimes \id$ are
injective. As a consequence of eq. \ref{Phis}, one has
$$(\bigoplus_{i}\Phi_{s_i})\hat{\cal K} = ({\cal K}\otimes \id)\Phi_s.$$

As stated above, ${\cal K}\otimes \id$ is injective; moreover,
$\bigoplus_{i}\Phi_{s_i}$ is bijective. Hence $\Phi_s$ is injective.
But since $\hat{\cal K}$ is also injective, and
$$(\bigoplus_{i}\Phi_{s_i})^{-1}({\cal K}\otimes \id)\Phi_s = \hat{\cal K},$$
it follows that $\Phi_s$ is bijective, completing the proof.
\end{proof}

\subsection{Quantum \v{C}ech cocycles and bundle reconstruction}
Now we introduce the concept of ``cocycle on a quantum space with values
in a quantum group''. This will play an important role in subsequent
developments, especially in the following reconstruction theorem and
in our main theorem. In fact --- it may be loosely said --- our definition
makes the cocycles take values in the classical subgroup of the quantum
group. That this should be the case could be foreseen on the basis of
lemma \ref{tau-lemma}.
\begin{defin}
Let $C(X)$ be a quantum space with covering $\{\kappa_i\}$,
$\kappa_i : C(X) \rightarrow C(U_i)$. Since $\kappa_i$ are surjective,
$C(U_i) \approx C(X)/\ker \kappa_i$. We shall denote
$C(U_{ij})\equiv C(X)/\{\ker \kappa_i + \ker \kappa_j\}$,
where $\{\ker \kappa_i + \ker \kappa_j \}$ is the smallest
two-sided closed $\star$-ideal containing $\ker \kappa_i$ and $\ker \kappa_j$.
Analogously, $C(U_{ijk}) \equiv C(X)/\{\ker \kappa_i + \ker \kappa_j
+ \ker \kappa_k\}$.
We shall also need the natural projections
$$\Pi^{ij}_k : C(U_{ij}) \rightarrow C(U_{ijk}).$$

A collection of homomorphisms $\tau_{ij} : C(G) \rightarrow Z(C(U_{ij}))$
will be called a $C(G)$-valued cocycle on $C(X)$ associated with
the covering $\{\kappa_i\}$ if the following conditions hold:
\begin{enumerate}
\item $\tau_{ii} = I_{C(U_{i})}\eps$
\item $\tau_{ji} = \tau_{ij}\circ S$
\item Defining the product
$$\tau_{ij}\ast \tau_{jk} : C(G) \rightarrow C(U_{ijk})$$
by
$$\tau_{ij}\ast \tau_{jk} = m_{C(U_{ijk})}(\Pi^{ij}_k \tau_{ij} \otimes
\Pi^{jk}_i \tau_{jk})\Delta,$$
require
$$\tau_{ij}\ast \tau_{jk} = \Pi^{ik}_j \tau_{ik}.$$
\end{enumerate}
\end{defin}
\begin{defin}
Two cocycles $\{\tau_{ij}\}$, $\{\tau_{ij}'\}$ associated with
the same covering $\{\kappa_i\}$ of $C(X)$ will be called equivalent
if, for some family of homomorphisms $\{\sigma_i\}$, $\sigma_i :
C(G) \rightarrow Z(C(U_i))$:
$$\tau_{ij}' = m_{C(U_{ij})}(m_{C(U_{ij})}\otimes \id)(\Pi^i_j
\otimes \id \otimes \Pi^j_i)(\sigma_i \otimes \tau_{ij} \otimes
\sigma_j S))(\id \otimes \Delta)\Delta,$$
where $\Pi^i_j : C(U_i) \rightarrow C(U_{ij})$ is the natural
homomorphism onto the quotient.
\end{defin}
{\em Note:} The above definition is correct, in spite of the occurence
of the quantum group antipode $S$ and the diagonal mapping
$m_{C(U_{ij})}$. This is due to the fact that both $\sigma_i$ and
$\tau_{ij}$ are required to be valued in the centers of the corresponding
algebras. In particular, this implies that they vanish on the kernel
of the classical projection of $C(G)$.
\begin{propo}
\label{cocycle}
Any quantum principal bundle $(C(P),C(X),G,\pi,\Gamma)$ together with
a set of local trivializations $(C(U_i),\kappa_i,\hat{\kappa_i})$
determines a cocycle $\{\tau_{ij}\}$ on $C(X)$ valued in $C(G)$,
associated with the covering $\{\kappa_i\}$.
\end{propo}
\begin{proof}
First, observe that if $\ker(\Pi^i_j \otimes \id)\hat{\kappa_i}=
\ker(\Pi^j_i \otimes \id)\hat{\kappa_j}$, one can define
$\Phi_{ij}: C(U_{ij})\otimes C(G) \rightarrow C(U_{ij})\otimes C(G)$
by the formula
$$\Phi_{ij} = [(\Pi^i_j \otimes \id)\hat{\kappa_i}]
[(\Pi^j_i \otimes \id)\hat{\kappa_j}]^{-1}$$
and $\Phi_{ij}$ thus defined will be a trivial principal bundle
automorphism. Moreover, $\Phi_{ij}\Phi_{ji} = \id$ and
$$[(\Pi^{ij}_k\otimes \id)\Phi_{ij}][(\Pi^{jk}_i\otimes \id)\Phi_{jk}]=
(\Pi^{ik}_j\otimes \id)\Phi_{ik}.$$

By lemma \ref{tau-lemma}, the corresponding $\{\tau_{ij}\}$
fulfill the conditions for being a $C(G)$-valued cocycle on $C(X)$
associated with $(\kappa_i, C(U_i))$.
\end{proof}

The initial assumption follows
from the subsequent lemma:
\begin{lemma}
A $G$-invariant two-sided ideal $i \subset C(U)\otimes C(G)$
is uniquely determined by $i \cap \pi(C(U)) = i \cap [C(U)\otimes I]$,
and is of the form $j\otimes C(G)$, where $j$ is a two-sided ideal
in $C(U)$.
\end{lemma}
\begin{proof}
According to theorem \ref{decomp},
$$i = \overline{\bigoplus_{\alpha \in \hat{G}} W_{\alpha}},$$
where $\hat{G}$ is the set of irreducible inequivalent representations
of $G$, and $W_{\alpha}$ are the corresponding invariant subspaces.
Since the action $\Gamma : i \rightarrow i \otimes C(G)$ is the
restriction of $\id \otimes \Delta$, any set of elements $f_j \in W_{\alpha}$
such that $\Gamma f_j = \sum_k f_k \otimes u^{(\alpha)}_{kj}$ must be
of the form $f_j = \sum_i h_i \otimes u^{(\alpha)}_{kj}, ~h_i \in C(U)$.
Since $i$ is an ideal in $C(U)\otimes C(G)$,
$$i\ni \sum_j(\sum_i h_i\otimes u^{(\alpha)}_{ij})(I\otimes S(u^{(\alpha)}))
= h_k \otimes I,$$
proving the claim.
\end{proof}

Another simple consequence of the above lemma is the following
corollary; essentially, it means that a $G$-invariant subset of
the bundle space $P$ is determined by its projection onto the
base space $X$.
\begin{coroll}
\label{inv-ideal}
Consider $\ker \hat{\kappa_i}$ --- an ideal in $C(P)$. For any $j$,
$\hat{\kappa_j}(\ker \hat{\kappa_i})$ is a $G$-right invariant ideal
in the product bundle $C(U_j)\otimes C(G)$, and is therefore determined
by its invariant elements, which belong to $\kappa_j(\ker \kappa_i)
\otimes I$. An ideal $i'$ in $C(P)$ with the property that
$\hat{\kappa_j}(i') = \kappa_j(\ker \kappa_i)\otimes C(G)$ for
all $j$ is the one generated by $\pi(\ker \kappa_i)$. But since
$\bigcap_j \ker \hat{\kappa_j} = \{0\}$, this ideal is unique.

By the same token, $\ker(\Pi^i_j \otimes id)\hat{\kappa_i}$ is
generated by $\pi(\ker \Pi^i_j\kappa_i)$. However,
$\ker(\Pi^i_j\kappa_i) = \{\ker \kappa_i + \ker \kappa_j\} =
\ker(\Pi^j_i\kappa_j)$. It thus follows that
$$\ker(\Pi^i_j\otimes id)\hat{\kappa_i} =
\ker(\Pi^j_i\otimes id)\hat{\kappa_j}.$$
\end{coroll}
Now, let $\{\hat{\kappa_i}\}$ and $\{\hat{\kappa_i}'\}$ be two
sets of local trivializations of a given bundle $(C(P),C(X),G,\pi,\Gamma)$
associated with the same covering $\{\kappa_i\}$ of the base space
$C(X)$. We state without proof the following
\begin{propo}
$\{\hat{\kappa_i}\}$ and $\{\hat{\kappa_i}'\}$ determine equivalent
cocycles $\{\tau_{ij}\}$ and $\{\tau_{ij}'\}$, respectively.
\end{propo}
It is moreover clear that two isomorphic principal bundles over the
same base space, supplied with local trivializations over the same
covering of the base space, also determine equivalent cocycles.

Now we proceed to consider the inverse problem: {\em i.e.} given
an equivalence class of cocycles, we will reconstruct the corresponding
quantum principal bundle, up to bundle isomorphism.
\begin{theorem}
\label{reconstruction}
Let $\{\tau_{ij}\}$ be a $G$-valued cocycle on the quantum space
$C(X)$, associated with the covering $(C(U_i), \kappa_i)$.
There exists a unique (up to isomorphism) quantum principal bundle
$(C(P), C(X), G, \pi, \Gamma)$ provided with a set of local
trivializations $\{\hat{\kappa_i}\}$, such that $\{\tau_{ij}\}$
is the corresponding cocycle.

For any cocycle $\{\tau_{ij}'\}$ equivalent to $\{\tau_{ij}\}$,
the corresponding bundle is the same (up to isomorphism) and the
corresponding $\{\hat{\kappa_i}'\}$ are such that $\hat{\kappa_i}' =
\hat{\kappa_i}\Phi_i$, where $\{\Phi_i\}$ are automorphisms of the
trivial bundles $C(U_i)\otimes C(G)$.
\end{theorem}
\begin{proof}
To construct a representative $C(P)$ of the isomorphism class of
bundles fulfilling the claim of the theorem, we apply the procedure
of connected sum of quantum spaces described in the Appendix.
We form the connected sum of trivial bundles $C(U_i)\otimes C(G)$;
the overlaps between the components of the sum are given by
$C(U_{ij})\otimes C(G)$, and the isomorphisms $\Phi_{ij}$ between
the overlaps are constructed from elements of the cocycle $\{\tau_{ij}\}$
following lemma \ref{tau-lemma}. The bundle structure on the disjoint
sum $\bigoplus_i (C(U_i)\otimes C(G))$ is given by its trivial bundle
structure, {\em i.e.} $\pi = \id \otimes I$ and $\Gamma = \id \otimes \Delta$.
It is easily verified that the connected sum is a $G$-invariant
subalgebra of the disjoint sum, and that it contains the image of $\pi$
restricted to $C(X)$ (understood as a connected sum of $C(U_i)$).
Finally, $\{\hat{\kappa_i}\}$ are given by the canonical projections
onto the components of the connected sum. It is also clear that the
cocycle determined by $C(P)$ and $\{\hat{\kappa_i}\}$ is again $\{\tau_{ij}\}$.

A cocycle $\{\tau_{ij}'\}$ which is equivalent to $\{\tau_{ij}\}$
determines isomorphisms $\Phi_{ij}': C(U_{ij})\otimes C(G) \rightarrow
C(U_{ij})\otimes C(G)$, which are given by $\Phi_{ij}' = \Phi_i \Phi_{ij}
\Phi_j^{-1}$, where $\Phi_i$ are the projections to $C(U_{ij})$ of
a family of automorphisms of $C(U_i)\otimes C(G)$. Together, the
automorphisms $\Phi_i$ determine an automorphism of the disjoint sum
$\bigoplus_i (C(U_i)\otimes C(G))$, under which $C(P)$ is taken to
$C(P')$. These two bundles are therefore isomorphic.
\end{proof}

\subsection{The classical sub-bundle}
In this subsection we will show that all the data of a quantum
principal bundle over $C(X)$ with structure group $G$ is actually
contained in its `classical sub-bundle', a bundle over $C(X)$
with structure group $G/$, the classical subgroup of $G$.

\begin{defin}
Let the quantum group $H = (C(H), v)$ be a subgroup of the quantum group
$G = (C(G), u)$, with $\rho : C(G) \rightarrow C(H)$ --- the subgroup
surjection. We will call $(C(Q), C(X), H, \pi_Q, \Gamma_Q)$
a sub-bundle of the principal bundle
$(C(P), C(X), G, \pi_P, \Gamma_P)$ with `co-embedding'
$\eta :C(P) \rightarrow C(Q)$, if $\eta$ is a surjective homomorphism,
and:
$$\eta \pi_P = \pi_Q,$$
$$(\eta \otimes \rho )\Gamma_P = \Gamma_Q \eta .$$
\end{defin}
\begin{propo}
\label{atlas}
Let $(C(Q), C(X), H, \pi_Q, \Gamma_Q)$ be a sub-bundle of the principal bundle
$(C(P), C(X), G, \pi_P, \Gamma_P)$. Given a set of local trivializations
$\{\hat{\lambda_i}\}$ of $C(Q)$, the formula
$$\hat{\kappa_i} = [((\id \otimes \eps_H)\hat{\lambda_i}\eta)\otimes \id]
\Gamma_P$$
provides a set of local trivializations of the bundle $C(P)$, over
the same covering of $C(X)$.
\end{propo}
\begin{proof}
The algebraic properties required for $\{\hat{\kappa_i}\}$ are easily
verified. The proof that they are surjective proceeds by techniques
analogous to those employed in Theorem \ref{sections}.
\end{proof}

\begin{theorem}
\label{subb}
Let $C(P)$ and $C(P')$ be two quantum principal bundles with the same
base space $C(X)$ and structure group $G$. If $C(P)$ and $C(P')$ both
have a sub-bundle $C(Q)$ over $C(X)$, with structure group $H$ being
a sub-group of $G$ under the same co-embedding $\rho: C(G) \rightarrow C(H)$,
they are isomorphic.
\end{theorem}
\begin{proof}
Using Proposition \ref{atlas} we construct local trivializations
$\{\hat{\kappa_i}\}, \{\hat{\kappa_i}'\}$ of $C(P)$ and $C(P')$,
respectively, in both cases over the same covering of $C(X)$. By
Proposition \ref{cocycle} these data provide two $G$-valued cocycles
on $C(X)$ associated with this covering. Obviously, these cocycles
are identical: they are obtained by composing the cocycle determined
by $C(Q)$ with the co-embedding $\rho$. By Theorem \ref{reconstruction},
the bundles $C(P)$ and $C(P')$ are therefore isomorphic.
\end{proof}
\begin{theorem}
\label{csb}
Let $(C(P), C(X), G, \pi, \Gamma)$ be a quantum principal bundle,
$\rho : C(G) \rightarrow C(G/)$ the homomorphism onto the classical
subgroup.

There exists a unique sub-bundle $(C(P/), C(X), G/, \pi /, \Gamma /)$
with co-embed\-ding $\eta : C(P) \rightarrow C(P/)$, which we will call
the {\em classical sub-bundle}.
\end{theorem}
\begin{proof}
Take an arbitrary set of local trivializations of $C(P)$,
$\{ \hat{\kappa_i}\}$, and define $\ker \eta = \bigcap_i \ker (\id \otimes
\rho )\hat{\kappa_i}$. $C(P/)$ is defined to be $C(P)/\ker \eta$,
with $\eta$ the natural projection onto the quotient.

Indeed, observe that for every $i$, $\ker (\id \otimes \rho)\hat{\kappa_i}$
is $G/$-invariant, since $\hat{\kappa_i}$ is $G$-covariant and
$\id \otimes \rho$ is $G/$-covariant. Hence $(\id \otimes \rho)\Gamma$
projects to a $G/$-action $\Gamma /$ on $C(P/)$. The atlas
$\{\hat{\kappa_i}\}$ on $C(P)$ defines an atlas $\{\hat{\kappa_i}/\}$
on $C(P/)$ by the formula
\beq
(\id \otimes \rho ) \hat{\kappa_i} = \hat{\kappa_i}/\circ \eta .
\eeq
To complete the proof we now proceed to show that:
\begin{enumerate}
\item{$\pi (C(X)) \cap \ker \eta = \{ 0\}$, thus $\pi / = \eta \pi$ is
injective, as required; }
\item{The subalgebra of $G/$-invariant elements in $C(P/)$ is equal to
$\pi /(C(X))$. }
\end{enumerate}
For the first point, observe that, by the formula relating $\hat{\kappa_i}$
and $\kappa_i$ (definition 7, b),
$$(\id \otimes \rho)\hat{\kappa_i}\pi x = \kappa_i(x) \otimes I_{C(G/)}$$
for all $i$ and for all $x \in C(X)$.
However, $\bigcap_i \ker \kappa_i = \{0\}$, proving the first claim above.

To prove the second claim: let $f \in C(P/)$ be $G/$-invariant,
{\em i.e.} $\Gamma / f = f \otimes I$. Consider now $\eta^{-1}f \subset C(P)$.
For any $f' \in \eta^{-1}f$,
$$g = (\id \otimes \mu_{G/})(\id \otimes \rho )\Gamma f' \in \eta^{-1}f,$$
where $\mu_{G/}$ is the Haar measure on $G/$, is $G/$-invariant, since
$\eta^{-1}f$ is $G/$-invariant and closed in $C(P)$. Thus for any $f$
we can take $f' \in \eta^{-1}f$ to be $G/$-invariant.

{}From $f'$ we obtain a collection $\{ f_i\} , f_i \in C(U_i)$, by taking
$f_i = (\id \otimes \eps )\hat{\kappa_i}f'$. It remains to be shown that
there exists an $h \in C(X)$ such that for all $i$, $f_i = \kappa_i h$.
If the latter holds, then obviously $f = \eta \pi h$: since $f'$ is
$G/$-invariant, $\hat{\kappa_i}f' \in C(U_i)\otimes C(G)$ is also
$G/$-invariant. As a consequence,
$$(\id \otimes \rho)\hat{\kappa_i}f' = [(\id \otimes \eps)\hat{\kappa_i}f']
\otimes I, $$
since for any $G/$-invariant $x \in C(G), \rho (x) = I \eps (x)$.

To show that such an $h \in C(X)$ exists, it is sufficient to prove that
for any pair $i, j$,
$$\Pi_j^i f_i = \Pi_i^j f_j.$$
To show this, we re-express the LHS by using the local automorphisms
$\Phi_{ij}$ introduced in the proof of Proposition \ref{cocycle} and
their expression in terms of the cocycle $\{ \tau_{ij}\}$, leading to
$$\Pi_{ij} f_i = m_{C(U_{ij})} (\Pi_i^j\otimes \tau_{ji})\hat{\kappa_j}f'.$$
Now, since $\ker \rho \subset \ker \tau_{ji}$, one may  write
$\tau_{ji} = \tau_{ji}' \rho$. Making use of the $G/$-invariance of $f'$,
we obtain
$$
\begin{array}{c}
m_{C(U_{ij})} (\Pi_i^j\otimes \tau_{ji})\hat{\kappa_j}f' =
m_{C(U_{ij})}(\Pi_i^j\otimes \tau_{ji}'(I_{C(G/)}\eps ))\hat{\kappa_j}f'= \\
 \Pi_i^j(\id \otimes \eps)\hat{\kappa_j}f'= \Pi_i^j f_j.
\end{array}
$$
This proves the claim.
\end {proof}

\begin{theorem}
\label{classic}
Let $(C(Q), C(X), H, \pi_Q, \Gamma_Q)$ be a quantum principal bundle
such that its structure group $H$ is the classical subgroup of a certain
quantum group $G$: $C(H) = C(G/)$. Then there exists a unique (up to
isomorphism) principal
bundle $(C(P), C(X), G, \pi_P, \Gamma_P)$ such that $C(Q)$ is its
classical sub-bundle.
\end{theorem}
\begin{proof}
The bundle $C(P)$ is obtained via the Reconstruction
Theorem (Th. \ref{reconstruction}). It suffices
to observe that the cocycle uniquely determined by $C(Q)$ and a set of
its local trivializations $(\hat{\kappa_i}^Q, C(U_i))$ extends uniquely
to a $G$-valued cocycle (over the same covering of $C(X)$):
$$\tau_{ij}^P : C(G) \rightarrow C(U_{ij}),$$
$$\tau_{ij}^P = \tau_{ij}^Q \rho ,$$
where $\rho$ is the canonical epimorphism $\rho : C(G) \rightarrow C(H)$.
It is obvious that the bundle reconstructed from $\{ \tau_{ij}^P\}$
fulfills the claim of the theorem.
\end{proof}
{\em Remark:}\/ The bundle $C(P)$ may also be reconstructed as
$$C(P) = \{f\in C(Q)\otimes C(G) : (\Gamma_Q \otimes \id -
\id \otimes (\rho \otimes \id )\Delta )f = 0 \},$$
with $\Gamma_P = \id \otimes \Delta \mid_{C(P)}$,
$\eta  : C(P) \rightarrow C(Q)$ given by $\eta = \id \otimes \eps \mid_{C(P)}$,
etc. We leave the proof as an exercise to the reader.
\section{Associated bundles}
In the classical situation, given a principal bundle $P$ with structure
group $G$, and a vector space $V$ which carries a representation of $G$,
it is standard to define a vector bundle associated to $P$ as a
suitable quotient space of $P \times V$. In this brief section we
limit ourselves to giving a quantum analog of this definition,
and stating and proving a simple proposition: a bundle associated to
a trivial principal bundle is itself trivial.
\begin{defin}
Let $V$ be a finite-dimensional linear space carrying a representation
$T$ of the quantum group $G$. A bundle associated to the principal bundle
$(C(P), C(X), G, \pi,\Gamma)$, corresponding to
the representation $T$, is the subspace
${\cal F}$ of $C(P))\otimes V$ determined by
$$ {\cal F} = \{ \alpha \in C(P)\otimes V:\: (\Gamma \otimes \id -
\id \otimes T)\alpha = 0 \}.$$
${\cal F}$ is naturally endowed with the structure of a left module
over $C(X)$: for $a \in C(X)$, $\alpha \in {\cal F}$,
$$a \cdot \alpha = \pi (a) \alpha.$$
\end{defin}
\begin{propo}
Let $C(P)$ be a trivial principal bundle. Then any associated bundle
${\cal F}$ is trivial, {\em i.e.} ${\cal F} \approx C(X) \otimes V$.
\end{propo}
\begin{proof}
$C(P)$, being trivial, may be identified (up to isomorphism) with
$C(X)\otimes C(G)$. Thus ${\cal F}$ is a subspace of $C(X)\otimes C(G)
\otimes V$. Note that $\id \otimes T: C(X) \otimes V \rightarrow
C(X)\otimes C(G)\otimes V$  is an injective mapping into
${\cal F} \subset C(X)\otimes C(G)\otimes V$. We will show that the
image of $\id \otimes T$ is actually equal to ${\cal F}$. Indeed,
for $\alpha \in {\cal F}$:
\begin{eqnarray*}
\lefteqn{\! \! \! \! \! \! \! \!
(\id \otimes T)(\id \otimes \eps \otimes \id )\alpha =
(\id \otimes \eps \otimes \id \otimes \id )(\id \otimes \id \otimes T)\alpha =}
\\ & & = (\id \otimes \eps \otimes \id \otimes \id )
(\id \otimes \Delta \otimes \id )\alpha = \alpha,
\end{eqnarray*}
proving the claim.
\end{proof}

\section{Examples}
In the present section we will give a few elementary examples of
quantum principal fiber bundles over non-commutative base spaces,
with the matrix quantum group $SU_q(2)$ as structure group. We begin by
recalling the definition of the quantum group $SU_q(2)$ \cite{woron1}:
\begin{defin}
Let $q \in [-1,1] \setminus \{ 0\}$. The quantum group $SU_q(2) = (A, u)$,
where $A$ is the universal $C^{\star}$ algebra generated by two
elements $\alpha,\; \gamma$ satisfying the relations
\beq
\begin{array}{ll}
\alpha^{\ast}\alpha + \gamma^{\ast}\gamma = I, &
\alpha \alpha^{\ast} + q^2 \gamma^{\ast}\gamma = I,\\
\gamma^{\ast}\gamma = \gamma \gamma^{\ast}, &
\alpha \gamma = q \gamma \alpha,\\
\alpha \gamma^{\ast} = q \gamma^{\ast} \alpha,& {}
\end {array}
\eeq
and
\beq
u = \left( \begin{array}{cc} \alpha , & -q\gamma^{\ast} \\
                            \gamma , & \alpha^{\ast}
\end{array} \right) .
\eeq
\end{defin}
For $q=1$ this reduces to the classical $SU(2)$ group. For $q \neq 1$,
the classical subgroup $SU_q(2)/$ is given by $SU_q(2)/ = (C(S^1), v)$,
where $S^1 = \{ e^{i\phi}: \phi \in {\bf R} \}$ and
$$ v = \left( \begin{array}{cc} \zeta, & 0 \\
                                0, & \overline{\zeta}
\end{array} \right) ,$$
with $\zeta \in C(S^1)$, $\zeta(e^{i\phi}) = e^{i\phi}$.
The co-embedding $\rho : SU_q(2) \rightarrow C(S^1)$ is given by
$\rho (\alpha) = \zeta ,\: \rho (\gamma) = 0$.

For our first example, we will consider $SU_q(2)$ principal bundles
over the quantum disk \cite{Klimek}. The quantum disk is defined
in the following:
\begin{defin}
Let $\mu \in (0, 1)$. The quantum disk $C(D_{\mu})$ is the universal
$C^{\star}$ algebra generated by the element $z$ satisfying the
relation
\beq
z z^{\ast} - z^{\ast} z = \mu (I - zz^{\ast})(I - z^{\ast}z).
\eeq
\end{defin}
In \cite{Klimek} it was shown that the algebra $C(D_{\mu})$ has the
following inequivalent irreducible $\star$-representations:
\begin{enumerate}
\item A family of one-dimensional representations (functionals)
defined by $z \mapsto e^{i\phi}$, with $\phi \in {\bf R}$. We see
that this family of functionals is parametrized by elements of $S^1$:
one may say that $S^1$ forms the classical boundary of the quantum
space $D_{\mu}$;
\item An infinite-dimensional representation $t$ defined as follows:
let $\cal H$ be the Hilbert space spanned by an orthonormal basis
$\{ e_n\}_{n \in {\bf N} }$. Then
\beq
t(z)e_n = \left\{ \begin{array}{ll}
                 0, & n = 0 \\
                 \sqrt{\frac{n\mu}{1 + n\mu} }e_{n - 1}, & n \geq 1;
\end{array} \right.
\eeq
\beq
t(z^{\ast})e_n = \sqrt{\frac{(n + 1)\mu}{1 + (n + 1)\mu} }e_{n+1}.
\eeq
\end{enumerate}
Furthermore (see Th. IV.7 of \cite{Klimek}), the algebra $C(D_{\mu})$ is
isomorphic to $C^{\star}(S)$, the unital $C^{\star}$ algebra generated
by the operator $S$ on $\cal H$, defined by $Se_n = e_{n+1}$ for all
$n \in {\bf N}$. This is true independently of the value of
$\mu \in (0, 1)$. In particular, it follows that the representation
$t$ is faithful.
The problem of classifying $SU_q(2)$ principal bundles over $D_{\mu}$
is solved by the following lemma:
\begin{lemma}
\label{coverings}
A $C^{\star}$ algebra $C(X)$ which admits a faithful irreducible
representation does not admit any nontrivial covering; that is, given
any covering $(\{\kappa_i \}, \{C(U_i)\} )_{i=1}^n$ of $C(X)$, for some $i$,
$\ker \kappa_i = \{ 0\}$.
\end{lemma}
\begin{proof}
Let us assume, to the contrary, that for all $i$, $\ker \kappa_i \neq \{0\}$.
Since $\bigcap_i \ker \kappa_i = \{0\}$, then for any collection
$\{ f_i \}_{i=1}^n$ such that $f_i \in \ker \kappa_i$, we have
$$f_n f_{n-1} \cdots f_2 f_1 = 0.$$
Since we have a faithful representation,
then (identifying elements of $C(X)$ with their images under this
representation) for any $f_1 \in \ker \kappa_1$ there exists an
$x_1 \in {\cal H}$ such that $f_1 x_1 \neq 0$. Note that, since $\ker \kappa_1$
is an ideal in $C(X)$, and the representation is irreducible, the image
of $x_1$ under the action of $\ker \kappa_1$ must form at least a dense
subset in $\cal H$ --- being a subspace of $\cal H$ invariant under
the action of $C(X)$. Now, for any $f_2 \in \ker \kappa_2$, there must
exist, in $\ker \kappa_1 x_1$, a vector $x_2$ such that $f_2 x_2 =
f_2 f_1 x_1 \neq 0$. Applying this argument repeatedly, we obtain
$$ f_n f_{n-1} \cdots f_2 f_1 \neq 0,$$
contradicting the assumption.
\end{proof}

{}From the above lemma we conclude that $D_{\mu}$ does not admit nontrivial
coverings. As a simple consequence, we can now state:
\begin{propo}
\label{disk-bundles}
All quantum principal fiber bundles over the quantum disk are trivial.
\end{propo}

As another example, we now consider a `quantum sphere', obtained by
gluing together two copies of the quantum disk $D_{\mu}$. This procedure
is an instance of the general construction of connected sum of
quantum spaces (see Appendix), and does not differ essentially from
gluing together two ordinary (classical) disks to form a {\em classical}
sphere $S^2$.
\begin{defin}
By quantum two-dimensional sphere $C(S^2_{\mu})$ \footnote{The quantum sphere
here defined in fact coincides, as a $C^{\star}$-algebra, with those
introduced by Podle\'s in \cite{podles2}. } we mean the subalgebra
of the direct sum $C(D_{\mu}) \oplus (D_{\mu})$ determined by the condition
$$
C(S^2_{\mu})=\{ f_1\oplus f_2 \in C(D_{\mu})\oplus C(D_{\mu}) :\;
\psi (f_1) = \psi (f_2)\; \mbox{\rm for all functionals $\psi$}  \}.
$$
\end{defin}
Equivalently, since (see above) the set of functionals on $C(D_{\mu})$
may be identified with $S^1$, we can introduce the
{\em classical projection}
$\rho : C(D_{\mu}) \rightarrow C(S^1)$ by
$$\rho (f)(\psi) = \psi (f);$$
then the condition above reads $\rho (f_1) = \rho (f_2)$.

The above construction provides automatically a non-trivial two-element
covering of $S^2_{\mu}$, given by
$$\kappa_{1,2}: C(S^2_{\mu})\rightarrow
C(D_{\mu}), \: \kappa_{1,2}(f_1 \oplus f_2) = f_{1,2}.$$
It is easily seen
that $\ker \kappa_1 \cap \ker \kappa_2 = \{0\}$, thus we indeed have
a covering. Observe now that
$$C(S^2_{\mu}) \supset \ker \kappa_1 =
\{ 0 \oplus f_2 : \rho (f_2) = 0 \},$$
and similarly for $\ker \kappa_2$;
thus $\{ \ker \kappa_1 + \ker \kappa_2 \}$ consists of elements of the
form $f_1 \oplus f_2$ such that $\rho (f_1) = \rho (f_2) = 0$. Therefore,
$$C(S^2_{\mu})/ \{\ker \kappa_1 + \ker \kappa_2 \} =
C(S^2_{\mu})/ \ker(\rho \oplus \rho ) = C(S^1).$$
One may thus say
that the two quantum disks intersect along an `equator', which is
an ordinary circle $S^1$.

According to theorem \ref{reconstruction}, principal bundles over the
base space $C(S^2_{\mu})$ may be reconstructed from a cocycle
$\tau : C(G) \rightarrow C(S^1)$. Following theorem \ref{classic},
$\tau$ is determined uniquely by a cocycle $\tau / : C(G/) \rightarrow
C(S^1)$, valued in the classical subgroup of $C(G)$. It follows that
the classification of the $G$ {\em quantum} principal bundles over
$C(S^2_{\mu})$ obtained as above is equivalent to that of {\em classical}
$G/$ principal bundles over $S^2$. In particular, $SU_q(2)$ bundles
over $C(S^2_{\mu})$ are classified by the integers (for $q \neq 1$).

In fact, the above classification of bundles over $C(S^2_{\mu})$ is
exhaustive: That {\em any} principal bundle with base space $C(S^2_{\mu})$
admits a set of trivializations over the covering we have been using
is a simple consequence of corollary \ref{inv-ideal} and
proposition \ref{disk-bundles}.

For the next example \footnote{This was also introduced in \cite{podles2}
as an example of a quantum sphere.}, consider the quantum space
obtained from $D_{\mu}$
by identifying all points of the boundary $S^1$: {\em i.e.}
the subalgebra of $C(D_{\mu})$ consisting of elements $f$ such that
$\rho (f) = c I \in C(S^1)$, $c \in {\bf C}$.
As shown by the analysis of \cite{podlesPhD}, this algebra has only
one nontrivial ideal, given by $\rho (f) = 0$; therefore, it does not
admit any nontrivial covering. As a consequence, all principal bundles
with this base space are trivial.

On the other hand, the above algebra is obtained in \cite{podles2}
as a quantum quotient space of $SU_q(2)$ by an action of its classical
subgroup $U(1)$. The quantum group $SU_q(2)$ thus displays part of the
features of a $U(1)$-principal bundle over the quotient (for $q =1$
this is the Hopf fibration of $S^3 \approx SU(2)$ over $S^2$).
It is not, however, a principal bundle according to our definition (for
$q \neq 1$), as it is not locally trivial.

A slight extension of the above examples may be obtained by removing
one or more non-intersecting disks from a classical sphere $S^2$
and gluing quantum disks onto the $S^1$ boundaries. The reader will
easily find that the classification of $G$-principal quantum bundles
over the base spaces thus obtained reduces to that of $G/$-principal
classical bundles over $S^2$. The corresponding remains true with
$S^2$ replaced by any compact two-dimensional surface.

It remains a challange to find interesting examples where quantum spaces
are glued together in an `essentially non-commutative' way.

\appendix
%
%
\section{Appendix}
Compact quantum spaces are a generalization of the notion of compact
topological spaces. It turns out that many basic notions of the
theory of compact topological spaces admit noncommutative extensions.
In this appendix, we review such extensions for those notions which
find essential application in the present paper: cartesian product,
closed subset, intersection and union of closed subsets, disjoint and
connected union of quantum spaces, covering of a quantum space by
closed subsets, and classical subset of a quantum space. All our
definitions are straightforward dualizations of the corresponding
definitions for point sets; we choose to summarize them here for the sake
of completeness and clarity.

We recall that within the approach adopted in the present paper,
a compact quantum space (a. k. a. noncommutative topological space)
is represented by a (separable) unital $C^{\star}$-algebra, in general not
commutative. In the commutative case, this $C^{\star}$-algebra
may be identified with the algebra $C(X)$ of continuous complex functions
on an ordinary compact topological space $X$\cite{GN}; we denote
by $C(X)$ the algebra corresponding to the quantum space $X$ also
in the noncommutative case. Mappings
between quantum spaces correspond to unital $C^{\star}$-homomorphisms;
in particular, homeomorphisms are represented by algebra isomorphisms.

Fiber bundles are a generalization of cartesian products of topological
spaces. For our purposes, a suitable extension of the notion of
cartesian product to quantum spaces is the following: consider the
$C^{\star}$-algebras $C(X_1)$ and $C(X_2)$. By a theorem
due to Gelfand, Naimark and Segal (see \cite{GN}),
every (separable) $C^{\star}$-algebra
has a faithful continuous representation in $B(H)$, the algebra
of bounded linear operators on a (separable) Hilbert space $H$.
Moreover, this representation is isometric (norm-preserving).
Thus $C(X_1)$ and $C(X_2)$ may be identified with certain closed subalgebras
of $B(H)$. Their algebraic tensor product is contained in
$B(H)\otimes B(H) \subset B(H\otimes H)$. Completing the image of
$C(X_1)\otimes C(X_2)$ with respect to the norm in $B(H\otimes H)$, we
obtain a separable $C^{\star}$-algebra, which we identify as the
tensor product of $C(X_1)$ with $C(X_2)$. It turns out that the resulting
algebra does not depend on the choice of (faithful) representations
of $C(X_1)$ and $C(X_2)$. The quantum space corresponding to this algebra
will be understood as the cartesian product of the quantum spaces
$X_1$ and $X_2$. For commutative algebras $C(X_1)$ and $C(X_2)$, the
above construction yields the algebra of continuous functions on
the (topological) cartesian product $X_1 \times X_2$.

By a closed subset $Y$ of the quantum space $X$ we mean the quantum space
represented by a $C^{\star}$-algebra $C(Y)$ obtained as a quotient
algebra of $C(X)$ by a (closed, two-sided) ${\star}$-ideal $i$. The natural
projection homomorphism of $C(X)$ onto the quotient plays the role of
the canonical embedding of $Y$ in $X$.

Given two closed subsets $Y_1$, $Y_2$ of the quantum space $X$, we
can define their intersection $Y_1 \cap Y_2 \subset X$ as the
quantum space corresponding to the quotient of $C(X)$ by the
minimal ideal containing both $i_1$ and $i_2$, the ideals involved
in constructing $Y_1$ and $Y_2$. Obviously this ideal $i_{12}$ is
unique and consists of elements of the form $f_1 + f_2$, $f_1 \in i_1$,
$f_2 \in i_2$. This allows us to treat $Y_1 \cap Y_2$ also as a subset
of $Y_1$ and $Y_2$.

Similarly, the union $Y_1 \cup Y_2 \subset X$ is represented by the
quotient of $C(X)$ by $i_1 \cap i_2$, and clearly contains $Y_1$,
$Y_2$ and $Y_1 \cap Y_2$ as closed subspaces. The reader will easily
find that the above definitions extend straightforwadly to arbitrary
finite families of closed subsets of a quantum space, and obey
the usual identities of set calculus.

Now we introduce a few notions enabling elementary surgery operations
on quantum spaces. The disjoint union of quantum spaces, $X_1 \cup X_2$,
is represented by the direct sum algebra $C(X_1) \oplus C(X_2)$, and
obviously contains $X_1$ and $X_2$ as closed subspaces, with
$X_1 \cap X_2 = \emptyset$ (the empty set is represented by the trivial
algebra $\{ 0 \}$). Given some additional data, one may furthermore
form connected sums of quantum spaces: take $Y_1 \subset X_1$ and
$Y_2 \subset X_2$, closed subsets of the respective quantum spaces;
provided $Y_1$ and $Y_2$ are isomorphic, a choice of this isomorphism
allows us to identify them, giving a connected union of $X_1$ with
$X_2$. The precise definition is as follows: let $C(Y_1) = C(X_1)/i_1$
and $C(Y_2) = C(X_2)/i_2$, and let $\Phi_{12} : C(Y_1) \rightarrow C(Y_2)$
be a given $C^{\star}$ algebra isomorphism. The connected union of
$X_1$ and $X_2$ corresponding to these data is defined to be
the quantum space represented by the $C^{\star}$ subalgebra of
$C(X_1) \oplus C(X_2)$ consisting of elements $(f_1,f_2)$ obeying
the condition
$$\Phi_{12}\Pi_1 f_1 = \Pi_2 f_2 ,$$
where $\Pi_{1,2}$ denote the respective embedding homomorphisms.
The above construction generalizes in a straightforward way to
finite families of quantum spaces, provided the corresponding
isomorphisms $\Phi_{ij}$ obey suitable consistency conditions, in exact
analogy with the case of ordinary topological spaces.
In the case when $X_1$ and $X_2$ are themselves given as closed subsets
of another quantum space, then with the natural choice for the required
data, their connected union coincides with their union as subsets.

A (finite) closed covering of a quantum space $X$ is a finite family
of closed subsets $U_i$ of $X$ such that $X = \bigcup_i U_i$.
In more detail, a covering of $X$ is given by a finite family
of algebras $C(U_i)$ and surjective homomorphisms
$\kappa_i : C(X) \rightarrow C(U_i)$, such that
$\bigcap_i \ker \kappa_i = \{ 0 \} $. It must be noted here that
existence of a nontrivial (finite closed) covering is a restrictive condition
on the quantum space $X$; by nontrivial covering we mean that
$\ker \kappa_i \neq \{ 0 \}$ for each $i$.

As the final notion, we now introduce the classical subset $X/$ of
the quantum space $X$. This is defined as the quantum space
represented by the quotient of $C(X)$ by its commutator ideal, {\em i.e.}
the smallest closed ${\star}$-ideal containing all elements of the
form $fg-gf$, for all $f, \: g \in C(X)$. We see that $C(X/)$ thus
obtained is a commutative $C^{\star}$ algebra; thus, by the Gelfand-Naimark
theorem, it may be identified with the algebra of continuous functions
on the set of its (linear and multiplicative) functionals. This set
carries the natural structure of a compact topological space, and
we identify it with $X/$.
\bigskip

{\em Acknowledgements:}\/ We wish to thank P. Podle\'s for helpful
discussions. We are also grateful to S.L. Woronowicz for his kind
interest, and to W. Pusz and S. Zakrzewski for useful comments.
This work is supported by KBN Grant 2 1047 91 01.

\end{document}